\documentclass[aps,
prl,
reprint,
graphicx
]{revtex4-1}

\usepackage{amsmath}
\usepackage{color}
\usepackage{graphicx}

\usepackage{natbib}
\usepackage{hyperref}

\begin{document}


\title{Atomic vapor spectroscopy in integrated photonic structures} 



\author{Ralf Ritter}
\affiliation{5. Physikalisches Institut and Center for Integrated Quantum Science and Technology, Universit\"at Stuttgart, Pfaffenwaldring 57, 70550 Stuttgart, Germany}
\author{Nico Gruhler}
\affiliation{Institute of Nanotechnology, Karlsruhe Institute of Technology, 76344 Eggenstein-Leopoldshafen, Germany}
\author{Wolfram Pernice}
\affiliation{Institute of Nanotechnology, Karlsruhe Institute of Technology, 76344 Eggenstein-Leopoldshafen, Germany}
\author{Harald K\"ubler}
\affiliation{5. Physikalisches Institut and Center for Integrated Quantum Science and Technology, Universit\"at Stuttgart, Pfaffenwaldring 57, 70550 Stuttgart, Germany}
\author{Tilman Pfau}
\affiliation{5. Physikalisches Institut and Center for Integrated Quantum Science and Technology, Universit\"at Stuttgart, Pfaffenwaldring 57, 70550 Stuttgart, Germany}
\author{Robert L\"ow}

\email{r.loew@physik.uni-stuttgart.de}
\homepage{www.pi5.uni-stuttgart.de}

\affiliation{5. Physikalisches Institut and Center for Integrated Quantum Science and Technology, Universit\"at Stuttgart, Pfaffenwaldring 57, 70550 Stuttgart, Germany}


\date{\today}

\begin{abstract}
We investigate an integrated optical chip immersed in atomic vapor providing several waveguide geometries for spectroscopy applications. The narrow-band transmission through a silicon nitride waveguide and interferometer is altered when the guided light is coupled to a vapor of rubidium atoms via the evanescent tail of the waveguide mode. We use grating couplers to couple between the waveguide mode and the radiating wave, which allow for addressing arbitrary coupling positions on the chip surface. The evanescent atom-light interaction can be numerically simulated and shows excellent agreement with our experimental data. This work demonstrates a next step towards miniaturization and integration of alkali atom spectroscopy and provides a platform for further fundamental studies of complex waveguide structures.    
\end{abstract}


\maketitle 

Over the past decades, alkali atoms were not only subject to a broad spectrum of research fields but also found their way into technological applications. Unlike solid state systems, the dispersion free properties of atoms are ideal for sensing and referencing tasks. While ultra cold atomic gases are best suited for ultra precise measurements and atomic clocks \cite{Ye2012,Muller2012,Croin2009,Bresson2013}, they usually require a large apparatus to do so. In contrast, devices based on thermal atomic vapors offer less precision, but have been successfully miniaturized and integrated for applications e.g. in magnetometry \cite{Weis2009,Walker2012,Sheng2013}, frequency referencing \cite{Schawlow1971,Millerioux1994}, or atomic clocks \cite{Knappe2004}. Several approaches exist to achieve atom-light interaction on a microscopic scale such as atoms in hollow core fibers \cite{Epple2014,Gaeta2010}, nanofibers \cite{Rauschenbeutel2011,Franson2010,Shariar2008}, micro- and nanocells \cite{Kuebler2010,Keaveney2012}, or anti-resonant reflecting optical waveguides (ARROW) on a chip \cite{Schmidt2007,Schmidt2010}. Recently, an important step towards integration was made by surrounding solid core optical waveguides on a chip by a rubidium vapor cladding \cite{Levy2013}, where the evanescent tail of the light mode interacts with the atomic vapor in close vicinity to the waveguide. With the existing technology of photonic integrated circuits on an industrial scale, this approach offers ideal conditions for using atomic vapors in e.g. sensing and communication applications. Light sources, interconnections, photonic devices and detectors can all be contained in a single chip, allowing for complex network and multiplexing designs, potentially even at the single photon level. Due to the small mode area of the evanescent field efficient atom light coupling is achieved and saturation can be reached already at low power. 
\begin{figure}[ht]
    \includegraphics[scale=1]{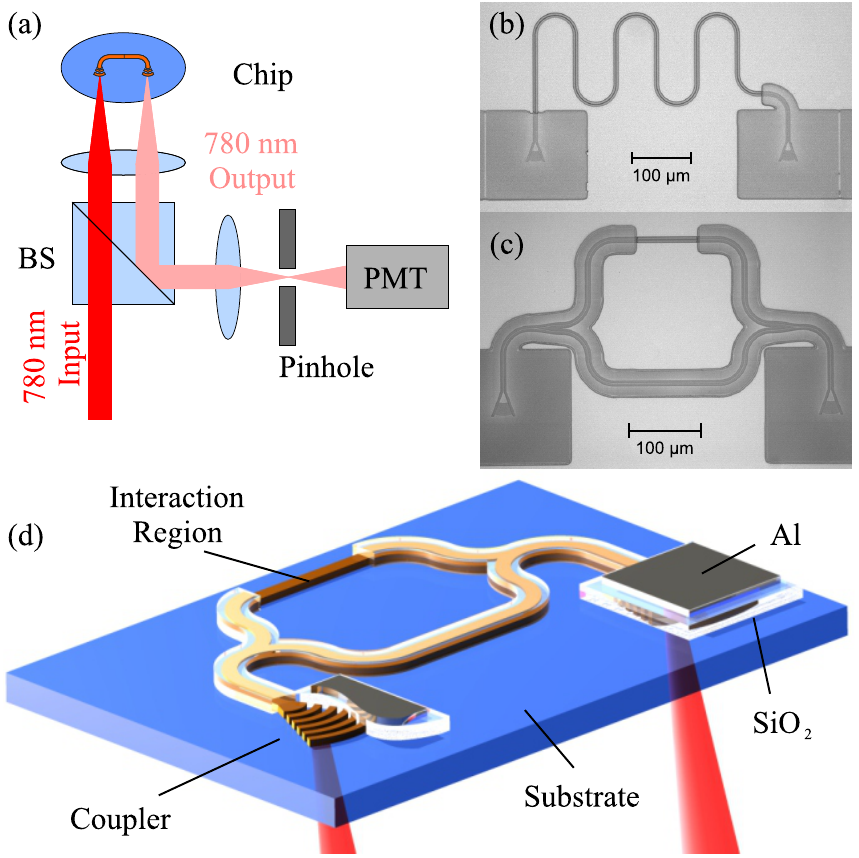}
    \caption{Experimental setup and layout of the structures. (a) Schematic of the experimental setup. For details see main text. (b) Microscopy image of the waveguide and (c) MZI structures. (d) Sketch of the MZI design and layer composition. The coverage of the front coupler is depicted sliced to reveal the grating coupler.}
    \label{fig:fig1}
\end{figure}  
In this letter we increase the level of complexity compared to previous works \cite{Levy2013} by adding Bragg couplers, curved waveguides and beam splitters to our silicon nitride photonic structures. We show evanescent atom-light coupling by means of a simple waveguide strip as well as a Mach Zehnder type interferometer (MZI), providing both transmission and phase shift information. Our experimental data can be well described by a theoretical model stemming from total internal reflection spectroscopy \cite{Nienhuis1988}.      

The substrate of our optical chip consists of a 4~mm thick 1.5~inch diameter fused silica vacuum window, covered with a 180~nm thick layer of silicon nitride (Si$_3$N$_4$). The photonic structures are created in this layer by electron-beam lithography and subsequent dry etching. Details on the fabrication process can be found in our previous paper\cite{Gruhler2013}. Focusing grating couplers are used for in- and out-coupling of light. This type of coupler can be placed arbitrarily on the two dimensional chip surface, therefore allowing a larger number of individual devices on a single chip and more flexibility as compared to e.g. butt coupling from the side of the chip where only a one dimensional distribution is possible. All devices are completely covered with a silicon oxide (SiO$_2$) layer, except for the regions where we want the atoms to interact with the light field. Additionally, we deposit a 100~nm thick opaque layer of aluminum on top of the SiO$_2$ layer above each grating coupler to avoid leakage of light through the coupler and the detection of fluorescence light from the atoms inside the chamber volume. This layer also acts as a local mirror and therefore increases the coupling efficiency. An overview of the layer composition is shown in figure~\ref{fig:fig1}(d). 
The chip is mounted into a custom made CF flange via a metal seal (Helicoflex) and connected to a UHV chamber with the structures facing the inside of the chamber. After pumping and baking the chamber to a pressure $<10^{-8}$~mbar, a rubidium ampule inside a bellow connected to the chamber is broken. We control the rubidium vapor pressure by the temperature of the bellow (reservoir), whereas we keep the chamber temperature always at a higher level (typically $\Delta T = 20$~K) to avoid condensation on the chip surface. 

In our experiments, we focus a 780~nm laser through the substrate onto the input grating coupler of a specific device (see figure~\ref{fig:fig1}(a)). The output coupler of this device is imaged onto a 100~$\rm \mu{}m$ pinhole to suppress any background light which is coming from the vicinity of this port. After the pinhole we detect the signal with a photo multiplier tube (PMT). For the chip used here we utilized standard couplers which are not yet optimized for mode matching between a focused Gaussian beam and the waveguide mode but rather for direct coupling from a fiber tip. With direct coupling a transmission of up to 6~\% is achieved for the presented structures, whereas a record coupling efficiency of -0.62~dB has been reported for grating couplers \cite{Berroth2014}. Using a focused Gaussian beam the overall efficiency is $2.5\times10^{-3}$ of the light which is sent to the chip and detected at the PMT for the simple waveguide. The waveguide quality was measured in the NIR and a propagation loss as low as 21~dB/m was obtained \cite{Gruhler2013}. In the visible regime the losses are usually slightly increased, however, they are still negligibly small especially compared to the absorption due to the atomic vapor.     

\begin{figure}[htb]
    \includegraphics[scale=1]{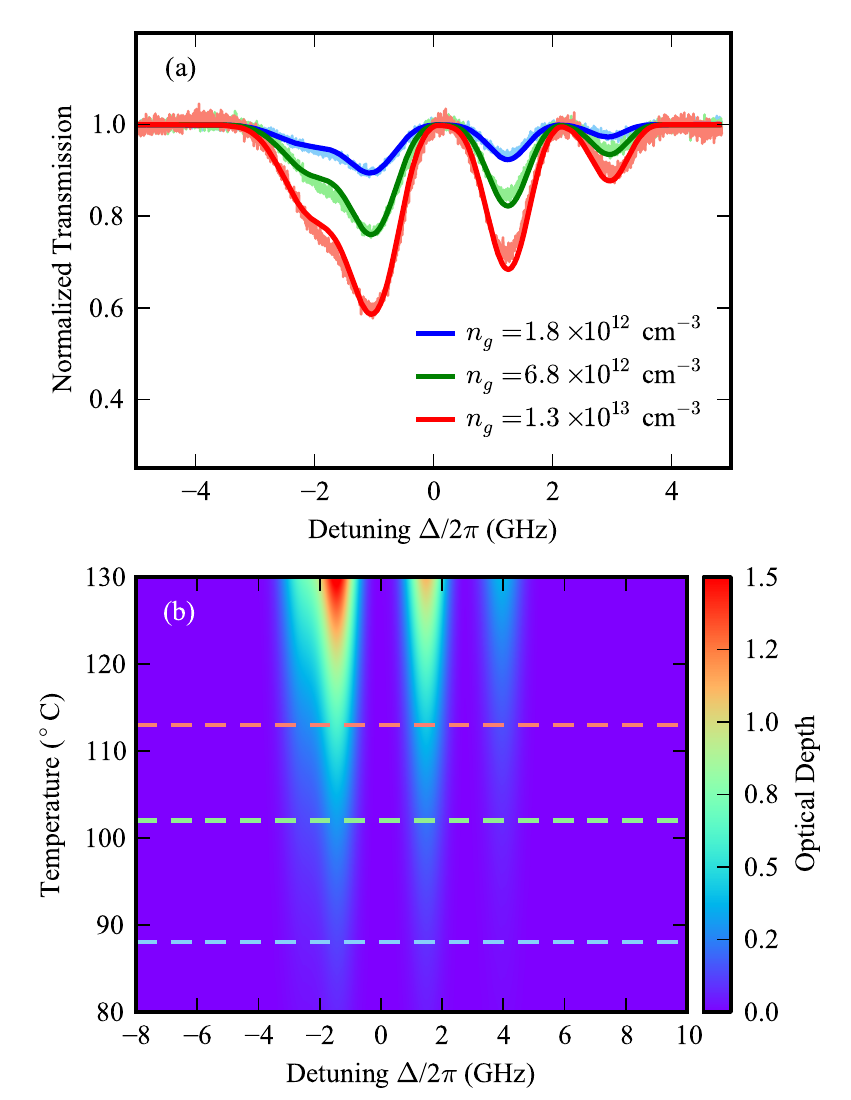}
    \caption{Absorptive features for the simple waveguide. (a) Absorption spectra of the Rb D$_2$ line for various atomic densities $n_g = 1.8\times10^{12}\textrm{cm}^{-3}$ (blue), $6.8\times10^{12}\textrm{cm}^{-3}$ (green), $1.3\times10^{13}\textrm{cm}^{-3}$ (red). The traces are normalized to the off-resonant transmission. The lines in darker color show the fit of the theory to the data. (b) Optical depth as a function of reservoir temperature and detuning calculated from our model. Dashed lines indicate the positions of the data from (a). At a temperature of $122^\circ$C an optical depth of 1 is achieved for the Rb$^{85}$ 5S$_{1/2}$, F$=3\rightarrow 5$P$_{3/2}$ transition.}
    \label{fig:fig2}
\end{figure}

\begin{figure*}[htb]
    \includegraphics[scale=1]{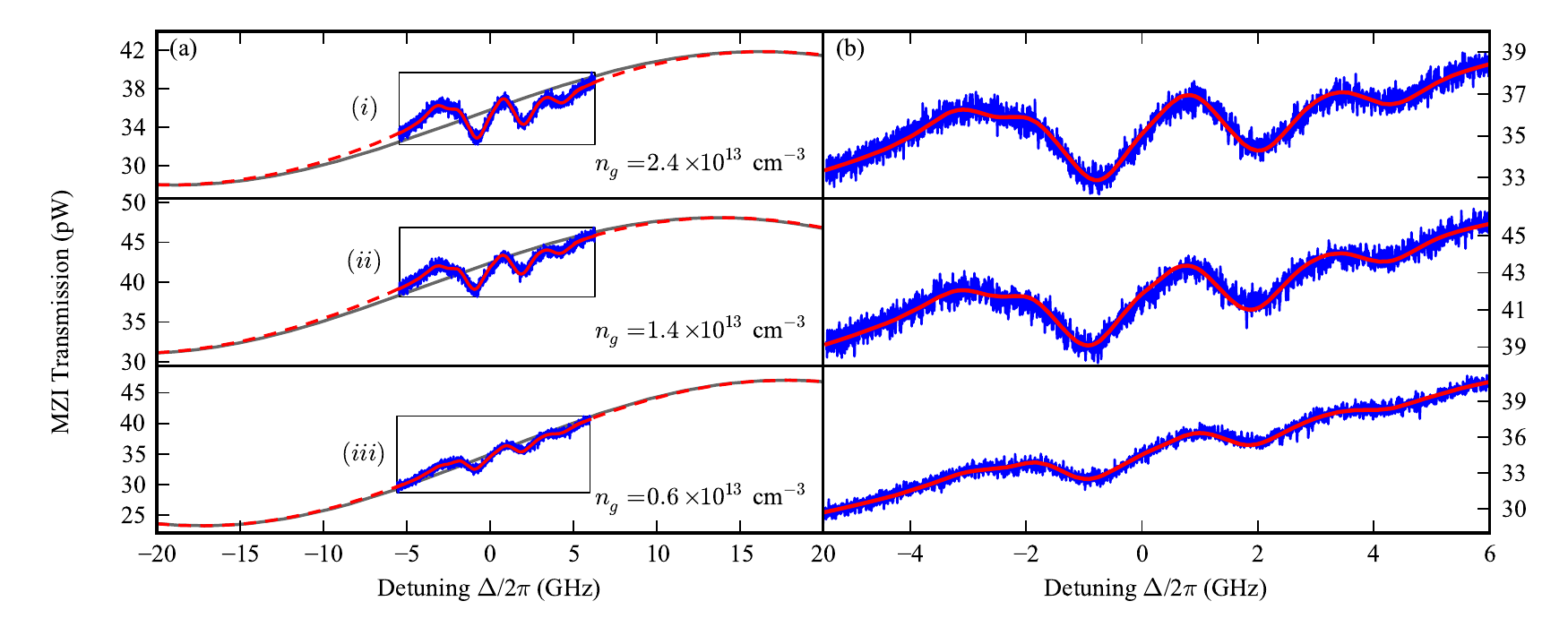}
    \caption{Transmission through a waveguide MZI with 2.2~mm path difference. (a) The gray solid lines show the calculated transmission of the MZI without any coupling to the atoms with a free spectral range of FSR $\approx$ 71~GHz. The insets contain the experimental traces (blue) for various atomic densities $n_g = 2.4\times10^{13}\textrm{cm}^{-3}$ ($i$), $1.4\times10^{13}\textrm{cm}^{-3}$ ($ii$), $0.6\times10^{13}\textrm{cm}^{-3}$ ($iii$). The solid red line is the theory fit while the red dashed line is the calculated transmission including the atomic contributions outside the scan range of the laser. Far off resonance this curve smoothly approaches the empty MZI fringe. (b) Zoom in on the insets from (a) with the experimental data (blue) and the theory fit (red).}
    \label{fig:fig3}
\end{figure*} 
For a first characterization of our system and to examine solely its absorptive properties, we investigated a simple waveguide as shown in figure~\ref{fig:fig1}(b) with a width of 1.1~$\rm \mu{}m$ and a height of 180~nm. By numerical simulation of the mode profile for this geometry, we infer that approximately 17\% of the TE mode interact with the atomic vapor, which is the preferably guided mode for our waveguide design. The uncovered length of the waveguide which is exposed to the atoms is $\sim$1.2~mm. In figure~\ref{fig:fig2}(a) we show the normalized transmission spectrum of such a device while scanning over the Rb D$_2$ line for different atomic densities. For a reservoir temperature of $113^\circ$C we already achieve an optical density of 0.5. In principle, higher densities can be achieved by simply increasing the reservoir temperature. However, as described later, we observed additional losses in the devices, possibly due to some deposition of Rb on the waveguide surface. Therefore we kept the reservoir temperature well below the chip temperature to reduce the amount of condensation. The line shape of the spectrum exhibits some distinct deviation from the well known conventional Rb D$_2$ spectrum. This is caused by the enhanced Doppler broadening due to the 1.6 times larger wave vector in the waveguide compared to free space propagation, and by the limited transit time of the atoms traveling through the evanescent field.      
     
To model our experimental data, we start with a finite-element analysis (COMSOL) to obtain the mode profile and propagation constant of the waveguide. With this information we can calculate the effective susceptibility $\chi_{\textrm{eff}}$ of the atoms surrounding the waveguide, as described in previous works \cite{Levy2013, Nienhuis1988, Guo1994}. In this calculation we neglect any saturation effects, but include Doppler broadening and transit time broadening as well as self broadening. Next, we add a cladding material with the complex refractive index $n_{\textrm{Rb}}=\sqrt{1+\chi_{\textrm{eff}}}$ to the waveguide in the COMSOL simulation. By running a frequency sweep with the detuning $\Delta$ around the center of the D$_2$ line, we obtain the complex propagation constant $\beta_{\textrm{Rb}}$. The transmission spectrum of the waveguide can then be calculated as
\begin{equation}
T = I_0 \times e^{-2\operatorname{Im}(\beta_{\textrm{Rb}}) L},
\label{eq:trans}
\end{equation}   
where $I_0$ is the intensity of the in-coupled light, and $L$ is the length of the interaction region. As shown in figure~\ref{fig:fig2}(a), this model fits well to our experimental data, where the frequency scaling, the center of the detuning and the Rb density are free fit parameters. In figure~\ref{fig:fig2}(b) we use our model to extrapolate the optical depth (O.D.) for a range of temperatures. From this follows that we can reach an O.D. of 1 at $122^\circ$C for the Rb$^{85}$ 5S$_{1/2}$, F$=3\rightarrow 5$P$_{3/2}$ transition and a waveguide with a length of 1.2~mm.

The second type of devices we investigated is a sub-mm Mach Zehnder interferometer as shown in figure~\ref{fig:fig1}(c) and (d). Here the in-coupled light is split by means of a 50/50 Y-branch into two arms with a path difference of 2.2~mm. The shorter arm (1) is completely covered with SiO$_2$, whereas the longer arm (2) is only partially covered, therefore offering a 2~mm long region for the guided mode to interact with the atoms. The modes from both arms are combined with a second Y-branch and guided to an output coupler. Besides the phase difference due to the unequal arm lengths, the light in the uncovered arm is picking up some additional phase caused by the real part of the susceptibility of the surrounding atoms. Also some of the light in the uncovered arm gets absorbed on resonance due to the imaginary part of the susceptibility. In total, this leads to a dispersive modulation of the MZI transmission as shown in figure~\ref{fig:fig3} for different atomic densities.       
   
The transmission of the MZI can be calculated as
\begin{equation}
T_{\textrm{MZI}} = \left|U_1 e^{i(\beta_1l_1+\phi_0)}+U_2 e^{i(\beta_1(l_2-l_{\textrm{I}})+\beta_{\textrm{Rb}}l_{\textrm{I}})}\right|^2,
\label{eq:MZI}
\end{equation}       
which describes the interference at the combining Y-branch. Here $U_1$, $l_1$ and $U_2$, $l_2$ are the light amplitudes and lengths of arm 1 and arm 2, respectively. The length of the interaction region is denoted by $l_{\textrm{I}}$. The propagation constant for a waveguide with SiO$_2$ cladding, as it is the case for arm 1, is denoted with $\beta_1$, whereas $\beta_{\textrm{Rb}}$ is the complex propagation constant for a Rb cladding, as described earlier. With $\phi_0$ we account for a phase offset due to a temperature dependent change of the arm lengths. A fit of this model to our data is also plotted in figure~\ref{fig:fig3}(b) and shows excellent agreement with the Rb density, the amplitudes and the phase offset as the only free fit parameters. Figure~\ref{fig:fig3}(a) shows the MZI transmission for a larger detuning range with and without contribution from the atoms. From the fitted curves we deduce, that the amplitude in arm 2 is approximately ten times smaller than the amplitude in arm 1 and decreasing during the course of the experiment, thus causing a smaller visibility than expected from a 50/50 beam splitter. We attribute this behavior to the larger length of arm 2 and therefore higher propagation losses on the one hand. On the other hand, it seems that some of the Rb atoms stick to the uncovered waveguide surface, increasing the losses in this arm additionally. A similar behavior was also found with the simple waveguide structures where we observed decreasing transmission over time.

The most intriguing feature of a Mach Zehnder interferometer is of course its ability to measure phase shifts.   
\begin{figure}[tb]
    \includegraphics[scale=1]{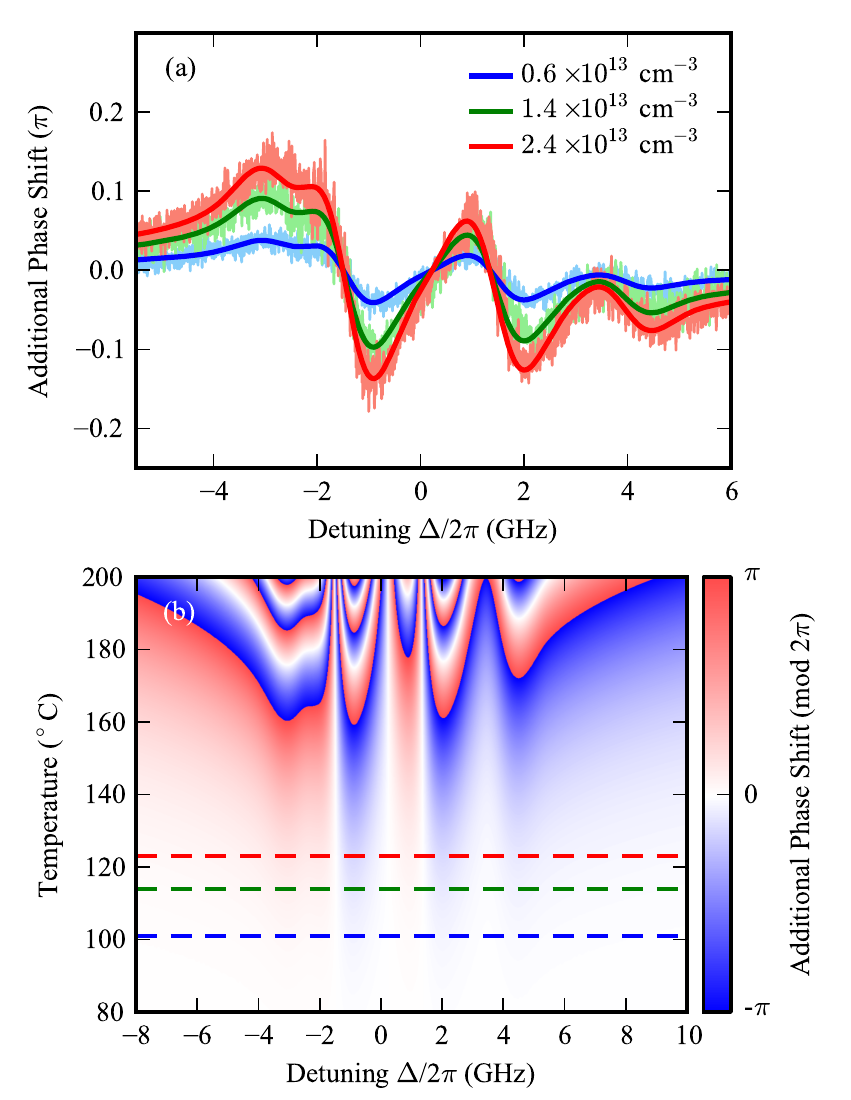}
    \caption{Additional phase shift in the MZI. (a) The bright traces show the phase shift extracted from the data for atomic densities $n_g = 2.4\times10^{13}\textrm{cm}^{-3}$ (red), $1.4\times10^{13}\textrm{cm}^{-3}$ (green), $0.6\times10^{13}\textrm{cm}^{-3}$ (blue). The dark curves are the corresponding calculated phase shifts for the parameters obtained from the fits in figure~\ref{fig:fig3}. (b) Calculated additional phase shift by the atoms modulo $2\pi$ as a function of detuning and reservoir temperature. Dashed lines indicate the positions of the data from (a).}
    \label{fig:fig4}
\end{figure}
We now can extract this phase shift $\Delta\varphi$ due to the presence of the atoms from our data transforming equation~\ref{eq:MZI} and subtracting the phase shift of the bare MZI:
\begin{eqnarray*}
\Delta\varphi&=&\cos^{-1}\left(\frac{T_{\textrm{MZI}}-\left|U_1\right|^2-\left|U_2\right|^2\exp\left(-2\operatorname{Im}(\beta_{\textrm{Rb}})l_{\textrm{I}}\right)}{2U_1U_2\exp\left(-\operatorname{Im}(\beta_{\textrm{Rb}})l_{\textrm{I}}\right)}\right)\nonumber\\
& &-\left[\beta_1l_1-(\beta_1(l_2-l_{\textrm{I}})+\beta_0l_{\textrm{I}})\right],
\label{eq:phase}
\end{eqnarray*}      
where $\beta_0$ is the propagation constant in the waveguide without cladding (vacuum). Figure~\ref{fig:fig4}(a) shows the phase shifts corresponding to the data in figure~\ref{fig:fig3}. For the data with the highest atomic density of $n_g = 2.4\times10^{13}\textrm{cm}^{-3}$, the light experiences an additional phase shift of up to $0.15\times\pi$. In figure~\ref{fig:fig4}(b) the calculated atomic phase shift for this particular device is shown in dependence of the reservoir temperature. Again we can extrapolate from our model, that an additional phase of $\pi$ is reached at a temperature of 160$^\circ$C, corresponding to an atomic density of $1.7\times10^{14}\textrm{cm}^{-3}$. Naturally a higher density is accompanied with strong absorption, but since the absolute value of the real part of the susceptibility is largest at the wings of the absorption lines, the off-resonant phase shift is still present, without much attenuation. The phase sensitivity of these devices can easily be increased by lengthening the uncovered arm or by decreasing the mode confinement.    

In conclusion, we have presented a hybrid system consisting of thermal alkali vapor and integrated photonic structures on a chip. Our optical chip houses several devices featuring grating couplers for flexible addressing of the individual devices. In future chip designs we are going to match the design of these couplers to our experimental conditions to improve the coupling efficiency. The transmission spectra of the simple strip waveguide revealed absorption of the evanescent tail for various atomic densities and showed line broadening due to the Doppler effect and the short transit time of the atoms through the evanescent field. To reach more narrow lines, two photon spectroscopy (e.g. electromagnetically induced transparency, EIT) can be utilized to cancel the Doppler shift, whereas the transit time broadening could be reduced by adding buffer gas. In addition to absorptive measurements we performed phase sensitive measurements using an integrated Mach Zehnder interferometer and could extract the atomic phase shift from our data for different Rb densities. By numerically simulating the light propagation in a waveguide surrounded by an atomic vapor cladding with a complex effective refractive index, we could reproduce the experimental data for both types of devices with excellent agreement.
Over time we witnessed some degradation of the waveguide structures, possibly due to some build up of a Rb layer on their surface. This process appears to be partially reversible: after we cool down the Rb reservoir but keep the chip at $\sim 200^\circ$C, the transmission increases again. For future chip generations we will investigate alternative materials and protective coatings to increase the lifetime of the devices. Additionally we aim to include the optical chips in a vapor cell using anodic bonding \cite{Daschner2014}, which is a further step towards integration and miniaturization and also allows for a better temperature control, therefore reducing the risk of Rb condensation on the waveguides. 
This hybrid system opens the door for future experiments with various waveguide geometries. The small mode area of the evanescent field on the order of $\lambda$ along the waveguide allows for systematic studies of interaction effects like self broadening in the one dimensional case, as it has been shown for two dimensions in a thin vapor cell \cite{Keaveney2012}. These interactions could also be utilized to add nonlinearity e.g. to a system of coupled ring resonators, which on their own create a synthetic gauge field for photons in the non-interacting regime \cite{Hafezi2013}.



%
%

%

\begin{acknowledgments}
We acknowledge support by the ERC under contract number 267100 and the Deutsche Forschungsgemeinschaft (DFG) with the project number LO1657/2. R.R. acknowledges funding from the Landesgraduiertenf\"orderung Baden-W\"urttemberg. 
\end{acknowledgments}


%

\end{document}